# BrainVoxGen: Deep-learning Framework for Synthesis of Brain Ultrasound to MRI

Shubham Singh[1], Mrunal Bewoor[2], Ammar Ranapurwala[2], Satyam Rai[2], Sheetal Patil[2]

[1] Department of Electronics and Computer Engineering, New York University

[2] Department of Computer Engineering, Bharati Vidyapeeth (Deemed to be) University

**Abstract**: The work proposes a novel deep-learning framework for the synthesis of three-dimensional MRI volumes from corresponding 3D ultrasound images of the brain, leveraging a modified iteration of the Pix2Pix Generative Adversarial Network (GAN) model. Addressing the formidable challenge of bridging the modality disparity between ultrasound and MRI, this research holds promise for transformative applications in medical diagnostics and treatment planning within the neuroimaging domain. While the findings reveal a discernible degree of similarity between the synthesized MRI volumes and anticipated outcomes, they fall short of practical deployment standards, primarily due to constraints associated with dataset scale and computational resources. The methodology yields MRI volumes with a satisfactory similarity score, establishing a foundational benchmark for subsequent investigations.

## 1. INTRODUCTION

Medical imaging is crucial in the diagnosis and treatment of diseases, providing insight into the information not available through the diagnosis through the naked eye [1]. There have been several modalities for the diagnosis of several specific anatomical segments, x-rays reveal most information about bone structure and density of objects while techniques like Ultrasound and MRI give structural insights into tissue imaging. Ultrasound is a cost-effective and widely accessible imaging modality offering non-invasive, real-time imaging, delivering high amounts of information with respect to blow flow and tissue condition but in low resolution [2]. While MRI is high resolution, it is expensive and potentially risks patients with metallic implants or pacemakers due to strong magnetic fields [3].

Over the last two decades, the advent of computing power has originated in the use of resources for enhancing and gaining better insight into the information available in the imaging than visible to human eye. The rise of Artificial intelligence has given birth to automation of the process of extracting information and developing insights, and deep learning notably performs well in this domain regarding this task.

This work utilizes deep learning techniques to convert ultrasound voxels into MRI voxels with the aim of extracting some extra information from the images [4]. The synthesized MRI images enhance diagnostic accuracy, enabling the potential for better treatment, especially with respect to the brain, which is one of the most complex and hard to diagnose and extract information from. This research is propelled by recent advancements in deep learning algorithms in image synthesis. Generative adversarial networks stand out as one such technique demonstrating success in image-to-image translation tasks [5] with higher accuracy compared to its predecessors.

## 2. PREVIOUS WORKS

### 2.1 Deep-learning in Medical Imaging Applications

Recent strides in cutting-edge Generative models have sparked intrigue in the space of 3D image synthesis using deep learning. A review of previous works in related topics helps understand the progress in the domain so far and inspires ideas for the work in the field.

Earlier work included a data-driven approach to synthesize CT images from corresponding MR images utilizing a 3D FCN structure and adversarial training. This method outperformed previous models in CT image prediction [6].

Another noteworthy work implemented a 3D cGAN for synthesizing FLAIR images from T1-weighted MRI. Through local adaptive fusion and a two-pathway 3D CNN model for brain tumor segmentation enhancing tumor segmentation compared to the baseline models [7].

Cross-modal attention techniques were explored to refine the synthesis of magnetic resonance images from unpaired ultrasound images. Leveraging nonlocal spatial information for multi-modal knowledge fusion and propagation, this approach demonstrated the potential for generating realistic MR images [8].

Additionally, transformer-based GANs were introduced for brain tumor segmentation in 3D MRI scans. This framework integrated a generator utilizing a 3D CNN-based encoder and decoder architecture, alongside a ResNet module and transformer block. Incorporating deep supervision and hierarchical feature analysis, this method showcased superior generalization for brain tumor segmentation compared to other approaches [9].

### 2.2 Image-to-image translation, 3D image generation

The Previous section of the work covered the application of deep learning in medical imaging, most of the work discussed involved a popular application and domain of research in the field, i.e. organ/ tissue segmentation. This section emphasizes the discussion on image-to-image translation, which in terms of medical imaging would be transforming images from one modality to another. Also covering a rapid albeit naive research in 3D image generation

For image-to-image translation, pix2pix is the pioneer model harnessing conditional adversarial networks for image-to-image translation tasks. Effective in tasks such as synthesizing photos from label maps and colorizing images [10]. Significant improvements in the model have been introduced since then such as pSp uses a StyleGAN that removes the adversary by encoding the features in the latent domain [11] and pix2pix-zero that makes training pix2pix models a zero-shot task, significantly increasing accuracy in the new images and speed of training the model [12].

The recent advancements in 3D-aware image synthesis have marked a significant milestone in the field, driven by the integration of generative visual models and neural rendering techniques. While these developments show promise, current approaches exhibit limitations in two critical aspects. Firstly, some methods lack an underlying 3D representation or rely on view-inconsistent rendering, leading to the synthesis of images that lack multi-view consistency. Secondly, certain techniques rely on representation network architectures that may not sufficiently capture the complexity of scenes, resulting in compromised image quality.

One notable contribution in this domain is the Periodic Implicit Generative Adversarial Networks. The pi-GAN model addresses the challenges by leveraging neural representations with periodic activation functions and volumetric rendering to generate view-consistent radiance fields. By incorporating these advancements, pi-GAN achieves state-of-the-art results in 3D-aware image synthesis across various real and synthetic datasets [13].

In comparison to existing methodologies, pi-GAN demonstrates superior performance in terms of both 3D representation and image quality. The utilization of periodic implicit generative networks enables pi-GAN to capture intricate scene details effectively, resulting in high-quality synthesized images that exhibit multi-view consistency.

GRAF, or Generative Radiance Fields, constitutes a pioneering model tailored for synthesizing radiance fields, exhibiting exceptional performance in the task of novel view synthesis within a single scene. Central to GRAF's efficacy is its ability to transcend the constraints posed by voxel-based representations, thereby facilitating finer control over both scene and camera attributes. Notably, the disentanglement achieved between camera and scene properties enhances reconstruction accuracy while effectively managing inherent ambiguities. Moreover, the introduction of a multi-scale patch-based discriminator further empowers GRAF to generate high-resolution images from unposed 2D inputs alone. Empirical validation substantiates the effectiveness of GRAF in generating 3D-consistent models with outstanding fidelity [14].

Deferred Neural Rendering introduces a novel paradigm by seamlessly integrating traditional graphics pipelines with learnable components. At the heart of this approach lies the concept of Neural Textures, learned feature maps integrated into the scene capture process atop conventional 3D mesh proxies. These Neural Textures encode a wealth of information surpassing that of traditional textures, enabling the synthesis of photo-realistic images even in the presence of imperfect original 3D content. Noteworthy is the explicit control afforded by the 3D representation, facilitating precise manipulation of generated outputs across a spectrum of applications, including video re-rendering and scene editing. Extensive experimentation underscores the superiority of Deferred Neural Rendering over conventional methods across diverse image synthesis tasks [15].

The use of CNNs, GANs, and Transformers in medical image generation and translation shows great promise [16]. Deep-learning techniques spotlighted in the section align to extract cross-modal information in medical imaging. However, none of them work with data as complex as brain images or with a task as complex as image-to-image translation.

## 3. METHODOLOGY

### 3.1. Data

The dataset comprising preoperative three-dimensional brain MRI and ultrasound data from twenty-three patients diagnosed with gliomas is available on the NIRD open-source dataset [17].

To prepare the data, tumor registrations that provide corresponding positions of the MRI images with respect to Ultrasound were used to align data in a one-to-one correspondence to each other such that each pixel in both images corresponds to the information provided in the image.

Post-alignment, the images were cropped to obtain the pixels that contained information such that both the images contained one-to-one corresponding information about each other.

The applied steps in the data preparation process can be summarized as follows:

1. Rotation and axial transformation to align each pixel in MRI to each pixel in Ultrasound using tumor registration as alignment references.
2. Once aligned in the correct axes, the images are cropped from all dimensions such that the bigger image is equal to the dimensions of the smaller image.

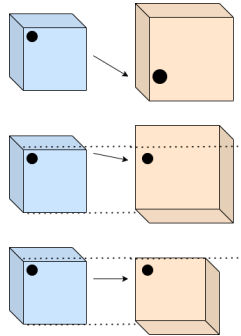

*Figure 1: Diagram illustrating the data processing process, rotation, and cropping*

The alignment and cropping procedure guaranteed the exact correspondence between MRI and ultrasound images, a requirement for the subsequent analytical phases of the research. The visual representations in Figures 2 and 3 highlight the data before and after of the process.

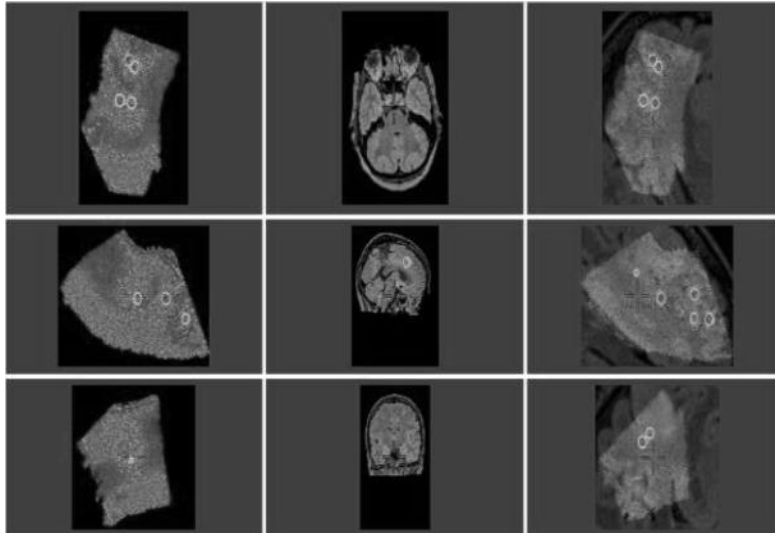

*Figure 2. NIRD three-dimensional dataset before preprocessing.*

*left to right: Ultrasound, MRI, Superimposed US-MRI*

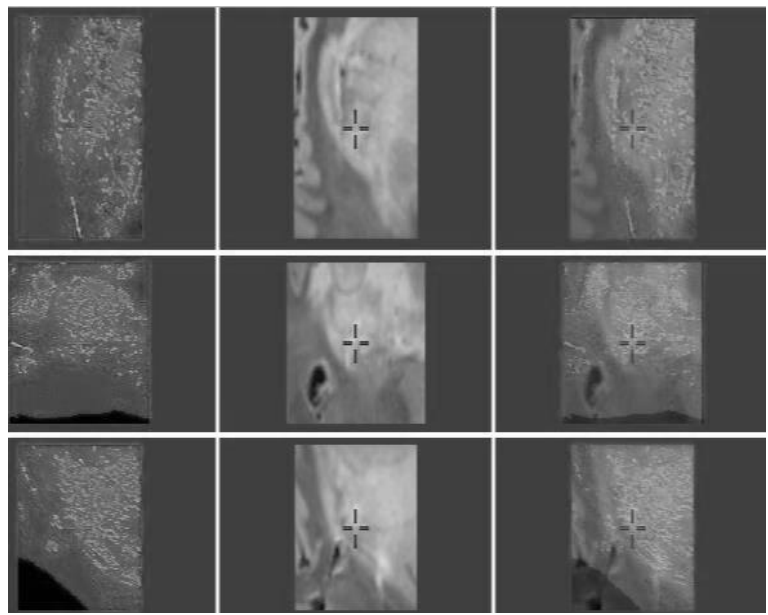

*Figure 3. NIRD three-dimensional dataset after preprocessing.*

*left to right: Ultrasound, MRI, Superimposed US-MRI*

### 3.2. Methodology

The work focuses on establishing the possibility of using deep learning models for generating three-dimensional volumes of brain images. Rather than using a complex state-of-the-art architecture like a Transformer, a GAN has been used to demonstrate the viability concept with plans to further extend this work to transformers and other state-of-the-art architecture. A modification of the Pix2Pix GAN architecture with multiple channels for the generation of images [18].

The proposed modification utilizes a three-dimensional ConvNet as the generator network, which takes a volume of ultrasound images as input and generates a corresponding 3D volume of MRI images as output. The generator network consists of multiple layers of three-dimensional up-sampled and down-sampled layers enabling the model to learn the underlying relationships between ultrasound and MRI features and generate MRI images from ultrasound images [19]. Meanwhile, the discriminator network is composed of multiple layers of three-dimensional convolutional and pooling layers followed by fully connected layers that map the features extracted from the input volumes to a final classification output.

The output of the discriminator is a scalar matrix that represents the probability that the input volume is real or fake [20]. Each iteration updates the generator image and discriminator matrix to decrease loss to increase similarity between ground truth and generated images. Figure 3. illustrates the architecture of the proposed model described in this section.

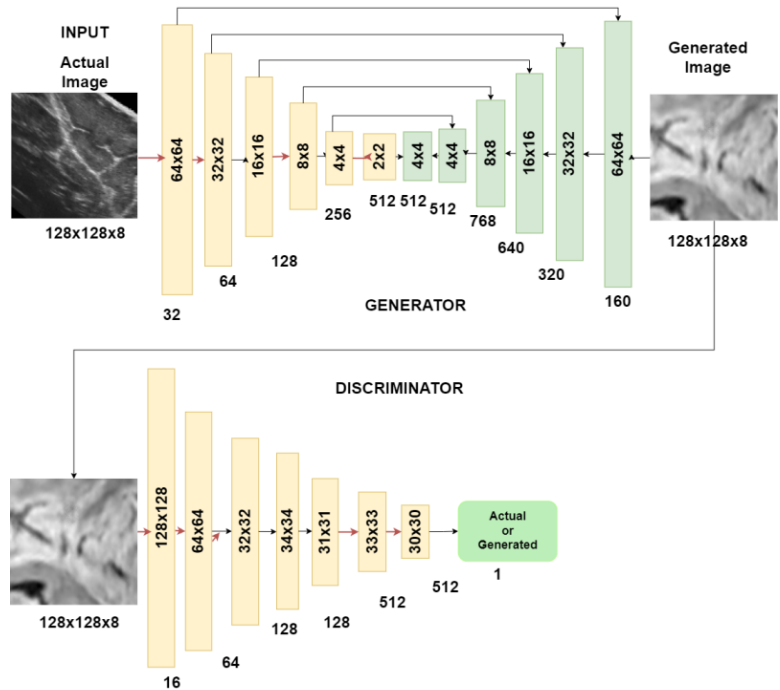

*Figure 4. Model Architecture*

**4. RESULTS**

The training spanned an extensive 70,000 epochs, utilizing partial slices of brain images as the primary input dataset. Central to the evaluation of the model's progress were two key metrics: Discriminator loss and Generator loss. These metrics, acting as fundamental indicators of the model's optimization trajectory, yielded valuable insights into the complex dynamics governing the training process [21].

Commencing the analysis, a thorough examination of the trajectory exhibited by the Discriminator loss throughout the training period is presented. Illustrated in Figure 4, the noticeable downward trend in this metric underscores the model's progressive capacity to discern genuine images from synthetic ones. However, the reduction of the loss curve notably slows down beyond the 40,000-iteration mark, suggesting a saturation point where further discernible improvements become increasingly challenging to achieve [22].

Conversely, Figure 6 depicts the fluctuating patterns characterizing the trajectory of the Generator loss. The initial steep decline in this metric highlights the model's rapid acquisition of salient data patterns. Subsequently, the loss curve demonstrates a propensity for fluctuation before ultimately converging to a plateau. This nuanced interplay of ascent and stabilization reflects the intricate dynamics between the generator's learning mechanisms and the evolving demands of the training regimen.

Further enhancing the analysis, Figures 7 and 8 provide a detailed examination of the L1 and Total Generator losses, respectively. The exponential decrease depicted by both metrics underscores the model's persistent pursuit of fidelity in generating voxel representations. Despite contending with the limitations imposed by a modest dataset comprising only eighteen voxels, painstakingly extrapolated to 288 voxels of dimensions 128x128x8, the achieved outcomes demonstrate a commendable degree of resilience.

However, Figure 9 brings attention to the synthesized images, revealing a notable disparity from the ground truth. While traces of feature extraction are discernible, the conspicuous lack of alignment underscores the need for further refinement.

In addition to qualitative assessments, Table 1 serves as a quantitative benchmark, outlining the evolution of Structural Similarity Index (SSIM) scores across discrete epochs of training. This established metric for evaluating image fidelity demonstrates a discernible increase in SSIM scores, indicative of the model's growing proficiency in generating images that approximate ground truth exemplars.

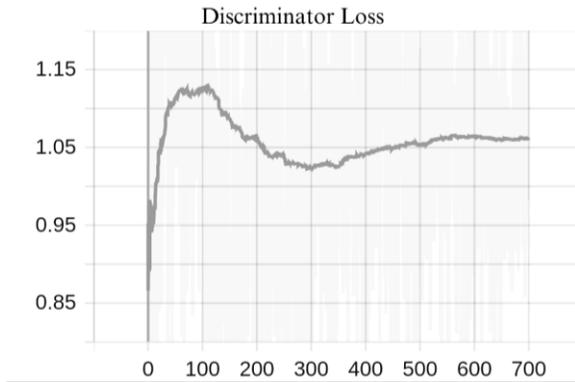

Figure 5. Discriminator loss

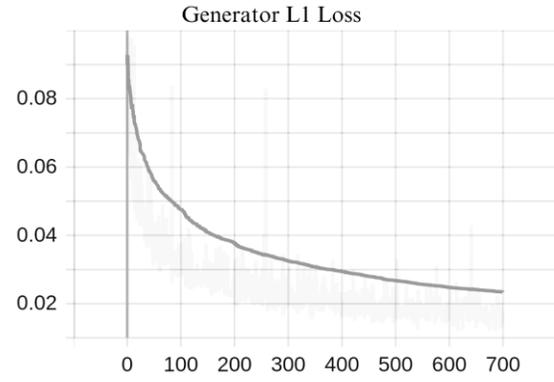

Figure 7. Generator L1 Loss

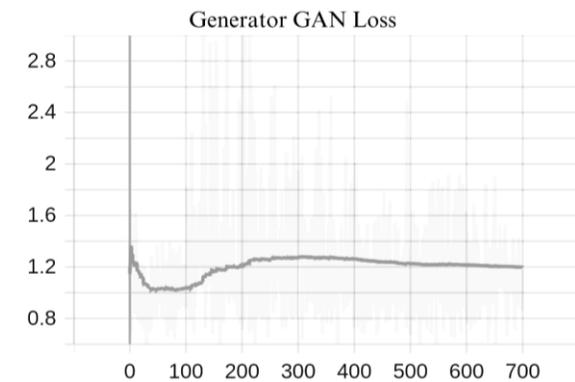

Figure 6. Generator GAN loss

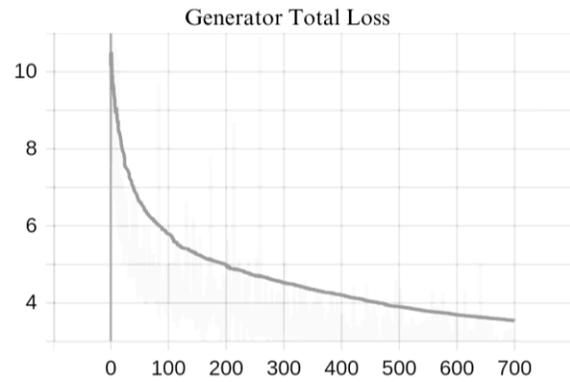

Figure 8. Generator Total Loss

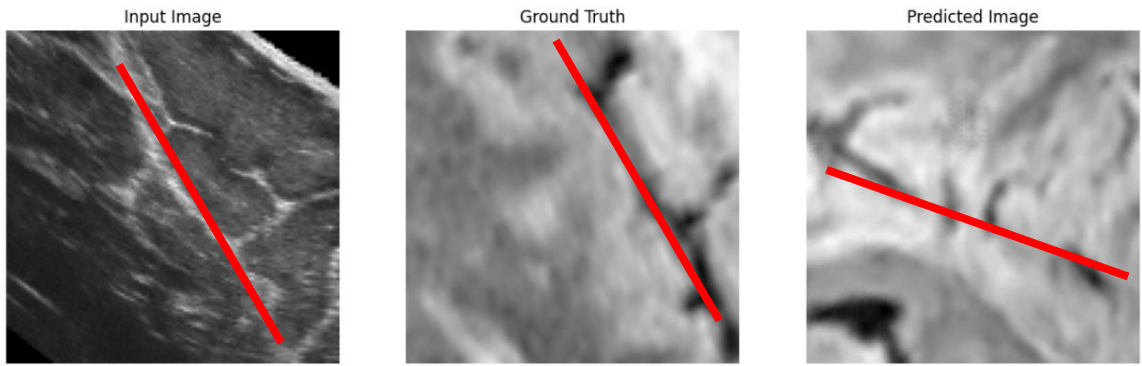

Figure 9: Input image, ground truth and predicted image

The results outlined in Table 1 are measured using the Structural Similarity Index, a recognized metric for evaluating image similarity [23].

*Table 1. Performance over training steps*

| Training Steps | SSIM Score | Generator Loss | Discriminator Loss |
|---|---|---|---|
| 10000 | 0.1032 | 5.87 | 1.15 |
| 20000 | 0.1077 | 5.02 | 1.05 |
| 30000 | 0.1793 | 4.61 | 1.03 |
| 40000 | 0.2146 | 4.24 | 1.04 |
| 50000 | 0.2589 | 3.94 | 1.05 |
| 60000 | 0.3183 | 3.78 | 1.06 |
| 70000 | 0.3623 | 3.47 | 1.05 |

## 5. DISCUSSION

The primary objective of this investigation was to establish a benchmark for subsequent works focused on the synthesis of MRI brain images utilizing ultrasound data. Employing sophisticated deep learning methodologies, particularly Generative Adversarial Networks (GANs), the study sheds light on the transformative potential of these techniques in distilling pertinent features from input imagery for translation tasks. The discerned efficacy of GANs in this context not only serves as a catalyst for further inquiry but also prompts meticulous optimization and refinement of these methodologies.

Moreover, the discourse advocates for an audacious exploration of unexplored research trajectories, emphasizing the importance of leveraging larger and more diverse datasets tailored explicitly for the intricate process of voxel-to-voxel translation in medical imaging. The conjecture posits that such expansive datasets hold the promise not only of facilitating a more nuanced training regimen but also of affording researchers the opportunity to delve into the intricacies inherent in the domain of medical image generation.

In conjunction with considerations regarding dataset dimensions, the integration of advanced architectural paradigms, notably the Transformer architecture, emerges as a potent strategy to enhance spatial understanding and information retention capabilities. The intrinsic capabilities of Transformer models in capturing prolonged dependencies and contextual nuances offer a tantalizing prospect for augmenting the fidelity and precision of synthesized images. By capitalizing on the latent potential of Transformer architectures, researchers are positioned to overcome one of the most challenging hurdles in medical image generation – the precise contextualization of adjacent tissues to bolster the overall information integrity embedded within the images.

The discourse underscores the role played by spatial information retention in the pursuit of precision in medical imaging. The adoption of the Transformer architecture heralds a noteworthy advancement in grappling with this challenge, laying the groundwork for theoretical conjectures and empirical explorations aimed at refining image generation processes. Furthermore, the scholarly discourse surrounding the assimilation of Transformer architectures in medical imaging applications portends to catalyze interdisciplinary dialogues and foster further progress in the field's advancement.

**6. CONCLUSIONS**

The proposed study embarks on a pioneering exploration into the application of generative adversarial networks (GANs) for the generation of voxel representations of medical imagery, with a specific focus on neuroimaging of the brain. Despite encountering inherent constraints within available datasets, this scholarly endeavor evinces a compelling potential to revolutionize the landscape of medical diagnostics, contingent upon rigorous refinement and optimization of the underlying methodologies.

Within the confines of voxel generation, the study confronts the pragmatic realities of data scarcity, encompassing challenges related to dataset size, diversity, and fidelity. Yet, amidst these challenges, it elucidates the latent efficacy of GANs in facilitating the creation of realistic voxel representations, promising enhanced capabilities in the detection and characterization of neurological anomalies and disorders. The realization of this potential necessitates a meticulous exploration of cutting-edge methodologies in deep learning. In this regard, the integration of large-scale models and transformative architectures, such as the Transformer, emerges as a salient avenue for further inquiry and refinement.

The incorporation of large-scale models offers the prospect of leveraging extensive datasets to bolster the robustness and generalization capacity of generated voxel representations. Concurrently, the integration of transformative architectures augments the fidelity and realism of synthesized voxel outputs by capturing intricate spatial dependencies and contextual nuances inherent within medical images. The pursuit of refinement extends beyond algorithmic development to encompass considerations pertaining to data acquisition and preprocessing methodologies. Efforts aimed at enhancing dataset quality, diversity, and representativeness are imperative for mitigating biases and fostering the generalizability of trained models.

While the proposed framework signifies a notable advancement in the application of GANs for voxel generation in medical imaging, its full potential hinges upon the persistent refinement and exploration of novel methodologies. Through a concerted interdisciplinary effort, the envisioned goal of transforming medical diagnostics through enhanced voxel generation may be realized, heralding a paradigm shift towards precision medicine and improved patient care.